\documentclass[11pt]{article}
\usepackage[hyperref]{acl2021}
\usepackage{times}
\usepackage{latexsym}

\usepackage{microtype}

\aclfinalcopy 

\usepackage{url}
\usepackage{amsmath, amssymb}

\usepackage{makecell} 

\title{Improving Neural Ranking Models with Traditional IR Methods}

\author{Anik Saha \\ Rensselaer Polytechnic Inst. \\ {\small \texttt{sahaa@rpi.edu}} \\ \And
  Oktie Hassanzadeh \\ IBM Research \\ {\small \texttt{hassanzadeh@us.ibm.com}}  \And 
  Alex Gittens \\ Rensselaer Polytechnic Inst. \\ {\small \texttt{gittea@rpi.edu}} \AND
  Jian Ni \\ IBM Research \\ {\small \texttt{nij@us.ibm.com}} \And
  Kavitha Srinivas \\ IBM Research \\ {\small \texttt{kavitha.srinivas@ibm.com}} \And 
  B{\"u}lent Yener \\ Rensselaer Polytechnic Inst. \\ {\small \texttt{yener@cs.rpi.edu}}
  }

\date{}

\begin{document}
\maketitle

\begin{abstract}
Neural ranking methods based on large transformer models have recently gained significant attention in the information retrieval community, and have been adopted by major commercial solutions.
Nevertheless, they are computationally expensive to create, and require a great deal of labeled data for specialized corpora. 
In this paper, we explore a low resource alternative which is a bag-of-embedding model for document retrieval and find that it is competitive with large transformer models fine tuned on information retrieval tasks.
Our results show that a simple combination of TF-IDF, a traditional keyword matching method, with a shallow embedding model provides a low cost path to compete well with the performance of complex neural ranking models on 3 datasets. 
Furthermore, adding TF-IDF measures improves the performance of large-scale fine tuned models on these tasks.
\end{abstract}

\section{Introduction}

Traditional information retrieval methods such as TF-IDF, and BM25 work very well for keyword based queries, but they are not as effective for natural language queries containing full sentences.
These models are based on the idea of exact match where tokens in the query have to be in a document to be considered relevant.
The relevance between documents is compared by the frequency of the matched tokens and their importance.
Although this works very well for keyword based queries, they are not as effective for natural language queries containing full sentences~\cite{DBLP:journals/ftir/GuptaB15}.

In recent years, a plethora of neural models~\cite{GuoNeuralRankingSurvey19} have been applied for ranking tasks.
These models generally represent a body of text like a query or an article with a low dimensional vector in the embedding space and measure their similarity based on their cosine distance.
Since neural models are trained to capture the meaning of a sentence or paragraph, they put more "attention" to the key words relevant to the semantics.
These models however will not work well on rare words because embeddings for low frequency words are not tuned.

We devise a novel retrieval method that combines a highly efficient neural retrieval model for conceptual retrieval along with keyword matching methods. 
Our method has the advantage of being weakly supervised, i.e., not requiring extensive training data or click-through data, and could scale to millions of documents. 
Our experiments over three large data sets with sentence-based queries show the effectiveness of our approach comparing with neural or traditional keyword-based matching methods alone.  We contrast performance on this model with transformer models fine tuned for information retrieval, to show that even in that case, performance improves with the addition of traditional IR metrics.

\section{Related Work}
Many neural ranking models have been proposed for IR tasks~\cite{GuoNeuralRankingSurvey19}, but none to our knowledge combine keyword-based techniques with neural techniques directly for relevance scoring.
DSSM uses a fully connected network for learning a semantic representation of the query and document, and uses that to rank search results.
The C-DSSM \cite{shen2014latent} model replaces the fully connected layer in DSSM with a convolutional layer. 
The motivation for this is to extract contextual features relevant to IR.
\citet{palangi2016deep} uses LSTM as an encoder in a similar fashion.
The DSSM, C-DSSM and LSTM-RNN models use click-through data from a search engine for creating the training set where the query is a normal web query and the document is the title of the clicked web page, so their focus is on the case where users have explicitly formulated queries relevant to the document.

Another set of models have been developed for retrieving semantically similar questions in online QA forums.  Note that the task here is matching questions that tend to be relatively short, and approximately the same length, as opposed to a conceptual query against a long document.
The BOW-CNN model \cite{dos2015learning} learns embeddings for matching similar questions in AskUbuntu and Quora questions datasets.
They train bag-of-words weights similar to TF-IDF and a neural feature extractor for learning representations of questions.
The recurrent convolutional network in \citet{lei2016semi} learns to weigh the words in a question title or body by using neural gates like LSTM for convolutional features.  They also pretrained an encoder-decoder model on the question and their bodies for learning better representation from the smaller training set. 
\citet{gillick2018end} converted this question retrieval task to a ranking task by creating a set of similar questions from pairs. 

Weak Supervision has been used to generate training samples from huge datasets for building neural models. 
\citet{dehghani2017neural} use a keyword-based ranker like BM25 as a first stage ranker to get a smaller collection (e.g. 1000 documents) and a neural model is used to rank them, thus limiting the results to keyword-based matches. 
Zamani et al. \shortcite{zamani2018neural} use weak supervision using BM25 to build training data for a neural model, but their approach then relies only on the neural model. 
Recent work by MacAvaney et al.~\shortcite{DBLP:conf/sigir/MacAvaneyYHF19} uses a content-based approach for creating training data. 
They use headlines and headings that are more similar to queries formulated by users for the retrieval.  In summary, we know of no work that directly tries to combine the advantages of keyword-based retrieval methods with neural methods in relevance scoring.

\section{Model}
\subsection{Bag-of-Embedding}
We train a dual encoder model for matching queries with articles. 
There are two identical encoders for queries and articles. 
This encoder module returns an embedding for a sequence of tokens by averaging their word embeddings. 
The similarity score for a pair of query and article is given by the cosine similarity.
\begin{equation}
    score(q,a) = \sigma(<v_q, v_a>)
\end{equation}

where $v_q$ and $v_a$ are normalized embeddings of query $q$ and article $a$ respectively.
Here, $\sigma$ is the sigmoid function. 

The model is trained with a margin loss function that maximizes the difference in score between a positive pair and a negative pair.
\begin{equation}
    L = \max(0, \delta - s_p + s_n)
    \label{eq:loss}
\end{equation}
where, $\delta$ is the margin, $s_p$ is a positive pair and $s_n$ is a negative pair of query and article. \\

The positive and negative examples are created from a batch of positive examples. 
Our dataset contains $1$ relevant article for each query.
So we treat all other articles in a batch as negative examples.
For a query, we select the article with the lowest similarity score in the batch to form the negative pair in Eqn. \ref{eq:loss}.

\subsection{Combination with TF-IDF}
We use a classic TF-IDF matching approach that transforms a sequence of tokens into a vector of the size of the vocabulary. Each element in this vector is the product of term frequency ($tf$) and inverse document frequency ($idf$), $\text{TF-IDF}(t,d) = tf(t,d) * idf(t)$ where $tf(t,d) = 1 + \log(f_{t,d})$, $f_{t,d}$ is the raw count of term $t$ in document $d$, and $idf(t) = \log\left(\frac{1 + n}{1 + df(t)}\right) + 1$, $n$ is the number of documents in the collection and $df(t)$ is the number of documents containing the term $t$.
Cosine similarity is used for the query and article vectors to rank them.
So, the ranking score is:
    $s_{\text{TF-IDF}}(q,a) = <v_q, v_a>$
where $v_q$ and $v_a$ are normalized TF-IDF vectors for query $q$ and article $a$ respectively. \\

We combine the embedding and TF-IDF models by aggregating their score for a pair of query and article.
\begin{equation}
    s(q,a) = s_{TF-IDF}(q,a) + s_{embed}(q,a)
\end{equation}
where $s_{embed}(q,a)$ is the dot product of the normalized embeddings vectors of $q$ and $a$.

\section{Experiments}
\subsection{Datasets}
We conducted experiments on 3 datasets to evaluate the ability of a model to semantically match a query and an article: 1) Signal Media News dataset \cite{Signal1M2016} containing $1M$ articles, 2) Wikipedia corpus containing $6M$ articles and 3) Google Natural Questions dataset with $300K$ question and article pairs.
We formed query-article pair from the news and Wikipedia dataset by selecting the first sentence of the article as query and treat the rest as the article.
In the natural questions dataset, we used the question as the query and the Wikipedia article containing the answer as the article. The news and wikipedia datasets were shuffled and split into train, validation and test sets. For natural questions, we use the validation set as the test set in our experiments since the test set is not public. Table~\ref{tab:data} shows the statistics. Table~\ref{tab:data_length} lists the average number of words in the queries and articles for these datasets.

\begin{table}[h]
    \small
    \centering
    \begin{tabular}{lccc} \hline
         Dataset & Train Size & Dev Size & Test Size \\ \hline
         News & 906.523K & 10K & 10K \\
         Wikipedia & 5.123M & 10K & 10K \\
         Natural Questions & 297.373K & 10K & 7.83K \\ \hline
    \end{tabular}
    \caption{Dataset statistics.}
    \vspace{-15pt}
    \label{tab:data}
\end{table}

\begin{table}[h]
    \centering
    \small
    \begin{tabular}{lcc} \hline
        Dataset & Query Length & Article Length \\ \hline
        News &  10 & 457 \\
        Wikipedia & 25 & 447 \\
        Natural Questions & 9 & 5643 \\ \hline
    \end{tabular}
    \caption{Query and article length}
    \label{tab:data_length}
    \vspace{-15pt}
\end{table}

\subsection{Training}
During preprocessing, we performed standard word-based tokenization.
Since we are interested in verbose queries, we did not include sentences shorter than 5 words as queries in the news and Wikipedia datasets.
Due to limited GPU memory, the articles were truncated to first 1000 tokens during training.
We trained the bag-of-embeddings model with Adam \cite{kingma2014adam} optimizer using a learning rate of 0.001.
The model was trained on the Wikipedia dataset for 20 epochs.
On both the news and natural questions datasets, the model was trained for 50 epochs.
Based on validation set performance, we set the embedding dimension to 768, the batch size to 1000 and the margin parameter $\delta$ to 0.5.
The code for experiments on the Signal Media News data set is available in \url{https://github.com/aniksh/dual_encoder}.

\subsection{Baselines}
\paragraph{TF-IDF.} We used both unigrams and bigrams for building the features in TF-IDF  after stopword removal. 
The \textit{scikit-learn} implementation is used for this method.

\paragraph{BM25.} Okapi BM25 \cite{DBLP:conf/trec/RobertsonWJHG94} is a strong baseline for information retrieval tasks. 
We tuned the parameters $k_1$ and $b$ of this model using 100 queries from the validation set.
We performed grid search with $k_1$ in the range $[0.5:0.5:5]$ and $b$ in the range $[0.3:0.1:0.9]$.

\paragraph{Dirichlet Language Model.} Language model with Dirichlet smoothing has been shown to be a strong baseline in verbose retrieval~\cite{DBLP:conf/cikm/PaikO14}. 
We tuned the smoothing parameter $\mu$ from the range \{100, 200, 300, 400, 500, 1000, 1500, 2000, 2500, 3000\}.

\paragraph{Sentence-Transformers. } We use the pretrained model from the sentence-transformers library \cite{DBLP:conf/emnlp/ReimersG19} for information retrieval, \texttt{msmarco-distibert-base-v2} \footnote{\url{https://www.sbert.net/docs/pretrained_models.html}}. 
This ranking model is based on BERT \cite{devlin2018bert} and was fine tuned on the MSMARCO passage ranking \cite{bajaj2016ms} dataset.


\subsection{Results}
We report the retrieval accuracy in the test set for 3 datasets using mean reciprocal rank (MRR) and mean precision at top-k (k = 1,3,10) results.
These are standard metrics for evaluating ranking models.
\[
MRR = \frac{1}{|Q|} \sum_{q\in Q} \frac{1}{r_q}
\]
Precision@k measures the existence of the relevant article in the top k predicted results from the model.
So precision@1 is equivalent to the accuracy of the ranking model.
The models are used to rank all articles in the test set for all queries.


In the News dataset, the TF-IDF + embedding model performs close to the BERT-based ranker (Table \ref{tab:news}.
MRR for the simple embedding model is 0.15 lower but it has a much higher precision at 3 and 10.

\begin{table}[h]
    \small
    \centering
    \begin{tabular}{l|c|c|c|c} \hline
    Model & MRR & \thead{Mean \\ P@1} & \thead{Mean \\ P@3} & \thead{Mean \\ P@10} \\ \hline
    TF-IDF & 66.21 & 58.34 & 71.23 & 80.32 \\
    BM25 & 67.63 & 61.86 & 71.2 & 77.81 \\
    LM-Dirichlet & 55.6 & 50.49 & 58.64 & 64.87 \\ \hline
    BERT-ranker & 76.59 & 71.58 & 80.13 & 84.96 \\
    TF-IDF + BERT-ranker & \textbf{82.96} & \textbf{79.15} & 85.76 & 89.05 \\ \hline
    TF-IDF + BOE & 82.81 & 77.5 & \textbf{87.22} & \textbf{91.13} \\ \hline
    \end{tabular}
    \caption{Retrieval performance on the News dataset.}
    \vspace{-10pt}
    \label{tab:news}
\end{table}

The bag-of-embedding model lags behind the BERT ranker on Wikipedia dataset (Table \ref{tab:wiki}).
Similar to the News dataset, we see that the bag-of-embedding model performs better in the precision@3 and precision@10 metrics. 
So, this model is not as accurate as the BERT-based model but it provides more correct results in the top 3 or the top 10 search results.

\begin{table}[h]
    \small
    \centering
    \begin{tabular}{l|c|c|c|c} \hline
    Model & MRR & \thead{Mean \\ P@1} & \thead{Mean \\ P@3} & \thead{Mean \\ P@10} \\ \hline
    TF-IDF & 66.62 & 58.96 & 72.43 & 79.88 \\
    BM25 & 67.75 & 63.97 & 69.93 & 74.23 \\
    LM-Dirichlet & 50.68 & 46.81 & 52.67 & 57.5 \\ \hline
    BERT-ranker & 77.99 & 74.55 & 79.76 & 84.07 \\
    TF-IDF + BERT-ranker & \textbf{80.45} & \textbf{77.4} & 81.93 & 85.71 \\ \hline
    TF-IDF + BOE & 79.33 & 71.99 & \textbf{84.59} & \textbf{92.1} \\ \hline
    \end{tabular}
    \caption{Retrieval performance on the Wikipedia dataset.}
    \vspace{-10pt}
    \label{tab:wiki}
\end{table}

In the Natural Questions dataset, the combination of TF-IDF and bag-of-embedding model outperforms the BERT ranking model on all metrics.

\begin{table}[h]
    \small
    \centering
    \begin{tabular}{l|c|c|c|c} \hline
    Model & MRR & \thead{Mean \\ P@1} & \thead{Mean \\ P@3} & \thead{Mean \\ P@10} \\ \hline
    TF-IDF & 70.07 & 59.65 & 77.34 & 88.59 \\
    BM25 & 71.48 & 61.09 & 79.30 & 89.17 \\
    LM-Dirichlet & 64.42 & 54.05 & 71.78 & 82.61 \\ \hline
    BERT-ranker & 73.84 & 68.66 & 84.94 & 92.4 \\
    TF-IDF + BERT-ranker & 78.69 & 68.68 & 87.31 & 94.62 \\ \hline
    TF-IDF + BOE & \textbf{79.66} & \textbf{69.68} & \textbf{88.14} & \textbf{95.40} \\ \hline
    \end{tabular}
    \caption{Retrieval performance on the Natural Questions dataset.}
    \vspace{-10pt}
    \label{tab:ques}
\end{table}

\subsection{Discussion}
It is evident from the results on these 3 datasets that the shallow embedding model performs better than the huge BERT-based retrieval model when the article length is high.
The average article length is close to 450 words for both the News and the Wikipedia datasets while it is more than 5000 for the Natural Questions dataset.
The complex BERT model was fine tuned on the MSMARCO passage ranking dataset where the average article length is about 55 words.
We know BERT uses the WordPiece tokenizer and has a maximum sequence length of 512 tokens.
So for long articles, there is a limitation to the quality of representation obtained by BERT.
Our embedding model was trained with a maximum sequence length of 1000 words.
So it can achieve better performance for matching queries to articles on the Natural Questions dataset.

\paragraph{Effect of TF-IDF} Adding TF-IDF scores to the relevance score from any embedding model improves its performance on all 3 datasets.
We show an example from the News dataset in Figure \ref{fig:example}. 
The BERT-ranker focused on the word \textit{spirit} to match with the top ranking article.
But the name \textit{Tom Wood} is an easy match for the actual article.
Since TF-IDF puts more weight on rare words for matching, adding the TF-IDF score to the prediction from BERT-ranker pushed this article to the top result.
This explains how TF-IDF can improve the performance of neural models with a negligible computational cost.

\begin{figure}[h]
    \centering
    \fbox{\parbox{\linewidth}{\small \textbf{Query}: \\
        Tom Wood : England need to recreate the spirit of London 2012 \\
        \textbf{Top result from BERT-ranker}\\
        Concept players , one of South Wales ’ most exciting theatre companies , return after the success of HMS Pinafore and the award winning a little night music with the blisteringly funny play blithe spirit . the company will perform the hilarious play at the paget rooms \dots\\
        \textbf{Top result after adding TF-IDF} \\
        Tom Wood insists England must inspire the type of patriotic pride that swept the nation during London 2012 to fuel their quest for World Cup glory . the eighth instalment of the global showpiece begins when Fiji visit Twickenham on Friday night \dots}}
    \caption{Example query from the News dataset and result from the BERT-based ranking model}
    \label{fig:example}
    \vspace{-10pt}
\end{figure}

\section{Conclusion}
In this paper, we showed that simple superposition of relevance scores from TF-IDF and neural ranking models can provide significant boost in retrieval performance.
We presented a scalable and efficient neural ranking model using bag of word embeddings, and showed its effectiveness through experiments on three datasets with different characteristics.
We plan to use this training framework to build more efficient neural ranking models from language models that can seamlessly work on longer articles.
A major challenge in evaluation of neural retrieval models is the lack of publicly available models and open-source implementation of the solutions. 
Hence, we intend to make our source code as well as our data sets publicly available to ensure the reproducibility of our results.

\bibliographystyle{acl_natbib}
\bibliography{ref}

\begin{thebibliography}{17}
\expandafter\ifx\csname natexlab\endcsname\relax\def\natexlab#1{#1}\fi

\bibitem[{Bajaj et~al.(2016)Bajaj, Campos, Craswell, Deng, Gao, Liu, Majumder,
  McNamara, Mitra, Nguyen et~al.}]{bajaj2016ms}
Payal Bajaj, Daniel Campos, Nick Craswell, Li~Deng, Jianfeng Gao, Xiaodong Liu,
  Rangan Majumder, Andrew McNamara, Bhaskar Mitra, Tri Nguyen, et~al. 2016.
\newblock Ms marco: A human generated machine reading comprehension dataset.
\newblock \emph{arXiv preprint arXiv:1611.09268}.

\bibitem[{Corney et~al.(2016)Corney, Albakour, Martinez, and
  Moussa}]{Signal1M2016}
David Corney, Dyaa Albakour, Miguel Martinez, and Samir Moussa. 2016.
\newblock \href {http://ceur-ws.org/Vol-1568/paper8.pdf} {What do a million
  news articles look like?}
\newblock In \emph{Proceedings of the First International Workshop on Recent
  Trends in News Information Retrieval co-located with 38th European Conference
  on Information Retrieval {(ECIR} 2016), Padua, Italy, March 20, 2016.}, pages
  42--47.

\bibitem[{Dehghani et~al.(2017)Dehghani, Zamani, Severyn, Kamps, and
  Croft}]{dehghani2017neural}
Mostafa Dehghani, Hamed Zamani, Aliaksei Severyn, Jaap Kamps, and W~Bruce
  Croft. 2017.
\newblock Neural ranking models with weak supervision.
\newblock In \emph{Proceedings of the 40th International ACM SIGIR Conference
  on Research and Development in Information Retrieval}, pages 65--74. ACM.

\bibitem[{Devlin et~al.(2018)Devlin, Chang, Lee, and
  Toutanova}]{devlin2018bert}
Jacob Devlin, Ming-Wei Chang, Kenton Lee, and Kristina Toutanova. 2018.
\newblock Bert: Pre-training of deep bidirectional transformers for language
  understanding.
\newblock \emph{arXiv preprint arXiv:1810.04805}.

\bibitem[{Dos~Santos et~al.(2015)Dos~Santos, Barbosa, Bogdanova, and
  Zadrozny}]{dos2015learning}
Cicero Dos~Santos, Luciano Barbosa, Dasha Bogdanova, and Bianca Zadrozny. 2015.
\newblock Learning hybrid representations to retrieve semantically equivalent
  questions.
\newblock In \emph{Proceedings of the 53rd Annual Meeting of the Association
  for Computational Linguistics and the 7th International Joint Conference on
  Natural Language Processing (Volume 2: Short Papers)}, pages 694--699.

\bibitem[{Gillick et~al.(2018)Gillick, Presta, and Tomar}]{gillick2018end}
Daniel Gillick, Alessandro Presta, and Gaurav~Singh Tomar. 2018.
\newblock End-to-end retrieval in continuous space.
\newblock \emph{arXiv preprint arXiv:1811.08008}.

\bibitem[{Guo et~al.(2019)Guo, Fan, Pang, Yang, Ai, Zamani, Wu, Croft, and
  Cheng}]{GuoNeuralRankingSurvey19}
Jiafeng Guo, Yixing Fan, Liang Pang, Liu Yang, Qingyao Ai, Hamed Zamani, Chen
  Wu, W.~Bruce Croft, and Xueqi Cheng. 2019.
\newblock \href {http://arxiv.org/abs/1903.06902} {A deep look into neural
  ranking models for information retrieval}.
\newblock \emph{CoRR}, abs/1903.06902.

\bibitem[{Gupta and Bendersky(2015)}]{DBLP:journals/ftir/GuptaB15}
Manish Gupta and Michael Bendersky. 2015.
\newblock \href {https://doi.org/10.1561/1500000050} {Information retrieval
  with verbose queries}.
\newblock \emph{Found. Trends Inf. Retr.}, 9(3-4):91--208.

\bibitem[{Kingma and Ba(2014)}]{kingma2014adam}
Diederik~P Kingma and Jimmy Ba. 2014.
\newblock Adam: A method for stochastic optimization.
\newblock \emph{arXiv preprint arXiv:1412.6980}.

\bibitem[{Lei et~al.(2016)Lei, Joshi, Barzilay, Jaakkola, Tymoshenko,
  Moschitti, and M{\`a}rquez}]{lei2016semi}
Tao Lei, Hrishikesh Joshi, Regina Barzilay, Tommi Jaakkola, Kateryna
  Tymoshenko, Alessandro Moschitti, and Llu{\'\i}s M{\`a}rquez. 2016.
\newblock Semi-supervised question retrieval with gated convolutions.
\newblock In \emph{Proceedings of the 2016 Conference of the North American
  Chapter of the Association for Computational Linguistics: Human Language
  Technologies}, pages 1279--1289.

\bibitem[{MacAvaney et~al.(2019)MacAvaney, Yates, Hui, and
  Frieder}]{DBLP:conf/sigir/MacAvaneyYHF19}
Sean MacAvaney, Andrew Yates, Kai Hui, and Ophir Frieder. 2019.
\newblock \href {https://doi.org/10.1145/3331184.3331316} {Content-based weak
  supervision for ad-hoc re-ranking}.
\newblock In \emph{Proceedings of the 42nd International {ACM} {SIGIR}
  Conference on Research and Development in Information Retrieval, {SIGIR}
  2019, Paris, France, July 21-25, 2019}, pages 993--996. {ACM}.

\bibitem[{Paik and Oard(2014)}]{DBLP:conf/cikm/PaikO14}
Jiaul~H. Paik and Douglas~W. Oard. 2014.
\newblock \href {https://doi.org/10.1145/2661829.2661957} {A fixed-point method
  for weighting terms in verbose informational queries}.
\newblock In \emph{Proceedings of the 23rd {ACM} International Conference on
  Conference on Information and Knowledge Management, {CIKM} 2014, Shanghai,
  China, November 3-7, 2014}, pages 131--140. {ACM}.

\bibitem[{Palangi et~al.(2016)Palangi, Deng, Shen, Gao, He, Chen, Song, and
  Ward}]{palangi2016deep}
Hamid Palangi, Li~Deng, Yelong Shen, Jianfeng Gao, Xiaodong He, Jianshu Chen,
  Xinying Song, and Rabab Ward. 2016.
\newblock Deep sentence embedding using long short-term memory networks:
  Analysis and application to information retrieval.
\newblock \emph{IEEE/ACM Transactions on Audio, Speech and Language Processing
  (TASLP)}, 24(4):694--707.

\bibitem[{Reimers and Gurevych(2019)}]{DBLP:conf/emnlp/ReimersG19}
Nils Reimers and Iryna Gurevych. 2019.
\newblock \href {https://doi.org/10.18653/v1/D19-1410} {Sentence-bert: Sentence
  embeddings using siamese bert-networks}.
\newblock In \emph{Proceedings of the 2019 Conference on Empirical Methods in
  Natural Language Processing and the 9th International Joint Conference on
  Natural Language Processing, {EMNLP-IJCNLP} 2019, Hong Kong, China, November
  3-7, 2019}, pages 3980--3990. Association for Computational Linguistics.

\bibitem[{Robertson et~al.(1994)Robertson, Walker, Jones, Hancock{-}Beaulieu,
  and Gatford}]{DBLP:conf/trec/RobertsonWJHG94}
Stephen~E. Robertson, Steve Walker, Susan Jones, Micheline Hancock{-}Beaulieu,
  and Mike Gatford. 1994.
\newblock \href {http://trec.nist.gov/pubs/trec3/papers/city.ps.gz} {Okapi at
  {TREC-3}}.
\newblock In \emph{Proceedings of The Third Text REtrieval Conference, {TREC}
  1994, Gaithersburg, Maryland, USA, November 2-4, 1994}, volume 500-225 of
  \emph{{NIST} Special Publication}, pages 109--126. National Institute of
  Standards and Technology {(NIST)}.

\bibitem[{Shen et~al.(2014)Shen, He, Gao, Deng, and Mesnil}]{shen2014latent}
Yelong Shen, Xiaodong He, Jianfeng Gao, Li~Deng, and Gr{\'e}goire Mesnil. 2014.
\newblock A latent semantic model with convolutional-pooling structure for
  information retrieval.
\newblock In \emph{Proceedings of the 23rd ACM international conference on
  conference on information and knowledge management}, pages 101--110. ACM.

\bibitem[{Zamani et~al.(2018)Zamani, Dehghani, Croft, Learned-Miller, and
  Kamps}]{zamani2018neural}
Hamed Zamani, Mostafa Dehghani, W~Bruce Croft, Erik Learned-Miller, and Jaap
  Kamps. 2018.
\newblock From neural re-ranking to neural ranking: Learning a sparse
  representation for inverted indexing.
\newblock In \emph{Proceedings of the 27th ACM International Conference on
  Information and Knowledge Management}, pages 497--506. ACM.

\end{thebibliography}

\end{document}